\documentclass[aps,prl,twocolumn,superscriptaddress]{revtex4-1}

\usepackage[sort&compress]{natbib}
\usepackage{graphicx}

\begin{document}

\title{Multiple time scale blinking in InAs quantum dot single-photon sources}

\author{Marcelo Davan\c co}\email{marcelo.davanco@nist.gov}
\affiliation{Center for Nanoscale Science and Technology, National Institute
of Standards and Technology, Gaithersburg, MD 20899,
USA}\affiliation{Maryland NanoCenter, University of Maryland, College Park,
MD}
\author{C. Stephen Hellberg}\email{steve.hellberg@nrl.navy.mil}
\affiliation{Center
for Computational Materials Science, Code 6390, Naval Research Laboratory,
Washington DC 20375}
\author{Serkan Ates}
\affiliation{Center for Nanoscale Science and Technology, National Institute
of Standards and Technology, Gaithersburg, MD 20899, USA}
\affiliation{Maryland NanoCenter, University of Maryland, College Park, MD}
\author{Antonio Badolato}
\affiliation{Department of Physics and Astronomy, University of
Rochester, Rochester, New York 14627, USA}
\author{Kartik Srinivasan}\email{kartik.srinivasan@nist.gov}
\affiliation{Center for Nanoscale Science and Technology, National
Institute of Standards and Technology, Gaithersburg, MD 20899, USA}
\date{\today}% It is always \today, today,

\date{\today}

\begin{abstract}
We use photon correlation measurements to study blinking in single,
epitaxially-grown self-assembled InAs quantum dots situated in circular Bragg
grating and microdisk cavities. The normalized second-order correlation
function $g^{(2)}(\tau)$ is studied across eleven orders of magnitude in
time, and shows signatures of blinking over timescales ranging from tens of nanoseconds to
tens of milliseconds. The $g^{(2)}(\tau)$ data is fit to a multi-level
system rate equation model that includes multiple non-radiating (dark) states, from which
radiative quantum yields significantly less than 1 are obtained. This behavior is observed even in
situations for which a direct histogramming analysis of the emission time-trace data
produces inconclusive results.
\end{abstract}

%\keywords{}


\maketitle

Single-photon sources based on single epitaxially-grown quantum dots (QDs) are
promising devices for photonic quantum information
science\cite{ref:Michler_book,ref:OBrien_Furusawa_Vuckovic}.  As single-photon source brightness
is crucial in many applications, III-V compound nanostructures like InAs
QDs in GaAs are of particular interest, both due to their short radiative lifetimes (typically $\approx1$~ns\cite{ref:Dalgarno_Warburton}) and
the availability of mature device fabrication technology for creating scalable nanophotonic structures
which can modify the QD radiative properties in desirable ways.
In particular, structures can be created to further increase the QD radiative
rate\cite{ref:Gerard1} and funnel a large fraction of the emitted photons
into a desired collection channel\cite{ref:Barnes2}.

The overall brightness of the source is, however, also influenced by the radiative efficiency of the QD,
which can deviate from unity for a variety of reasons, including
coupling of the radiative transition to non-fluorescing states.  Such
fluorescence intermittency, also called blinking, is an apparently ubiquitous
phenomenon in solid-state quantum
emitters\cite{ref:Nirmal_blinking,ref:Kuno_Nesbitt,ref:Stefani_PT_blinking,ref:Frantsuzov_Marcus_blinking_review},
being particularly pronounced in single nanocrystal QDs and organic molecules.
In contrast, blinking in epitaxially-grown III-V QDs has not received as much attention, largely due to the fact that such QDs, grown in ultra-high-vacuum
environments and embedded tens of nanometers below exposed surfaces,
typically do not express pronounced fluorescence intermittency~\cite{ref:Lounis_Orrit_SPS_review,ref:Michler_book_2009}.
Obvious blinking (at the $\approx 100$~ms to $\approx 1$~s time scales) has only been observed in InGaAs QDs
grown close to crystal defects\cite{ref:Wang_Shih_blinking} and in some InP
QDs\cite{ref:Pistol_Samuelson_PRB,ref:Sugisaki}. Sub-microsecond blinking
dynamics in InAs QDs have also been studied\cite{ref:Santori4}.

Here, we study blinking in InAs/GaAs QDs embedded in photonic nanostructures
that enable high collection efficiencies ($\approx$10~$\%$)\cite{ref:Davanco_BE,ref:Ates_JSTQE}.
As these devices do not exhibit pronounced fluorescence variations, we use photon
correlation measurements\cite{ref:Lippitz_Orrit_blinking_review} as a more
informative approach to investigate blinking over time scales ranging from
tens of nanoseconds to hundreds of milliseconds. The data is fit with a
multi-level rate equation model for the QD that yields estimates for the
transition rates and occupancies of the QD states, enabling an overall
estimate of the QD radiative efficiency. Notably, we find quantum yields
significantly less than 1 even in QDs which show no blinking in histogramming
analysis. This information is valuable in quantifying the extraction
efficiency of QD emission, and in understanding the ultimate brightness
achievable for QD single-photon sources. We anticipate that the importance of
this topic is likely to grow as epitaxially-grown QDs are incorporated within
photonic nanostructures with critical dimensions of tens of nanometers, at
which point surfaces play an important role\cite{wang.APL.3423}.

\begin{figure}[h]
\centerline{\includegraphics[width=8.5 cm]{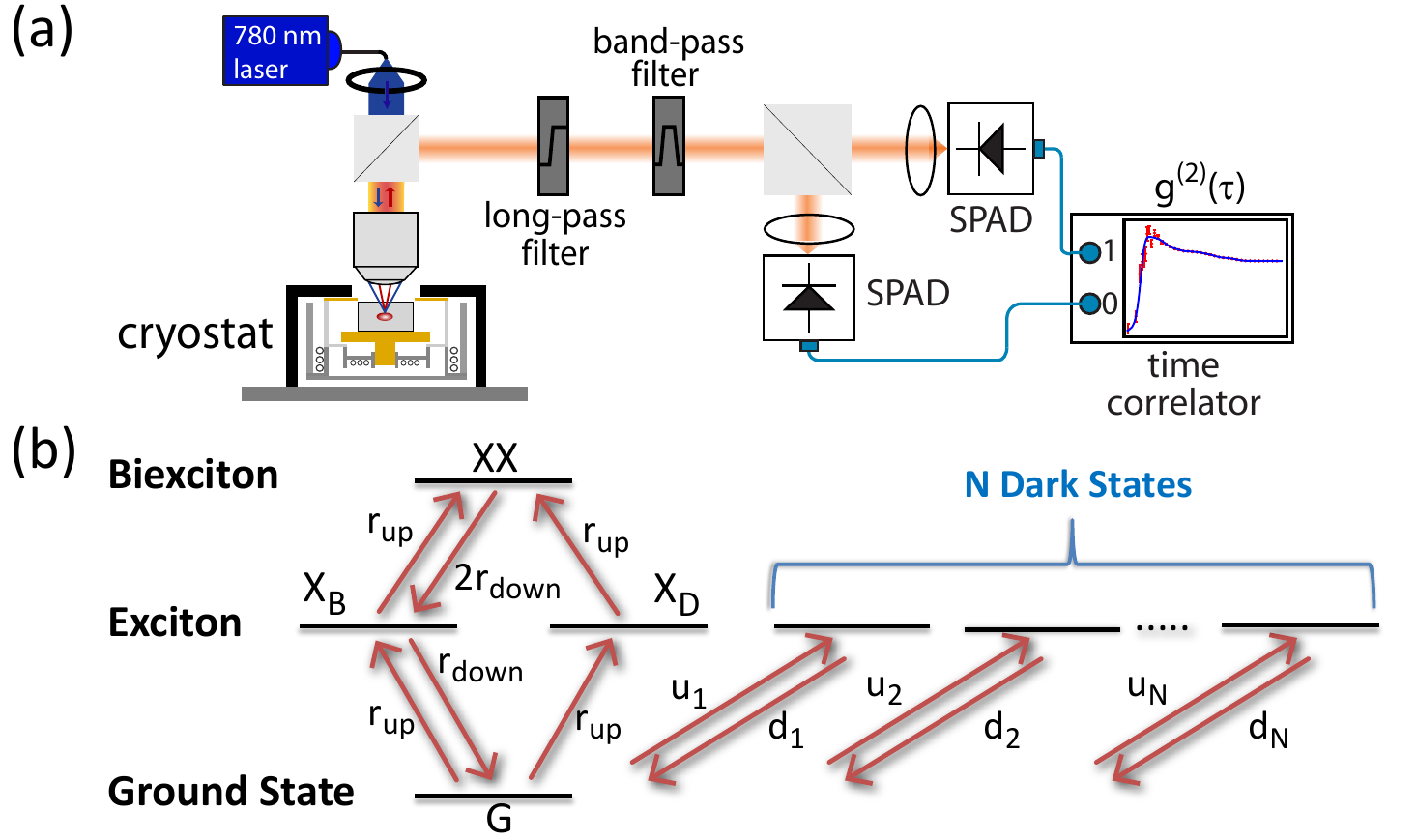}} \caption{(a)
Experimental setup (b) Multi-level model with $N$ dark states used to describe the QD behavior.
Pumping into the "bright" and "dark" single excitonic states ($\text{X}_{\text{B}}$ and
$\text{X}_{\text{D}}$) from the ground state ($\text{G}$),
and the biexciton state ($\text{XX}$) from the single exciton states, occurs at a rate
$r_{\text{up}}$. Spontaneous emission from the biexciton (exciton)
state occurs at a rate $2r_{\text{down}}$ ($r_{\text{down}}$).  The
up-transition (down-transition) rates between the ground state and dark states
are labeled $u_{\text{i}}$ ($d_{\text{i}}$), where $i$=1..$N$.}
\label{fig:Fig1}
\end{figure}

\begin{figure*}[ht]
\centerline{\includegraphics[width=17 cm]{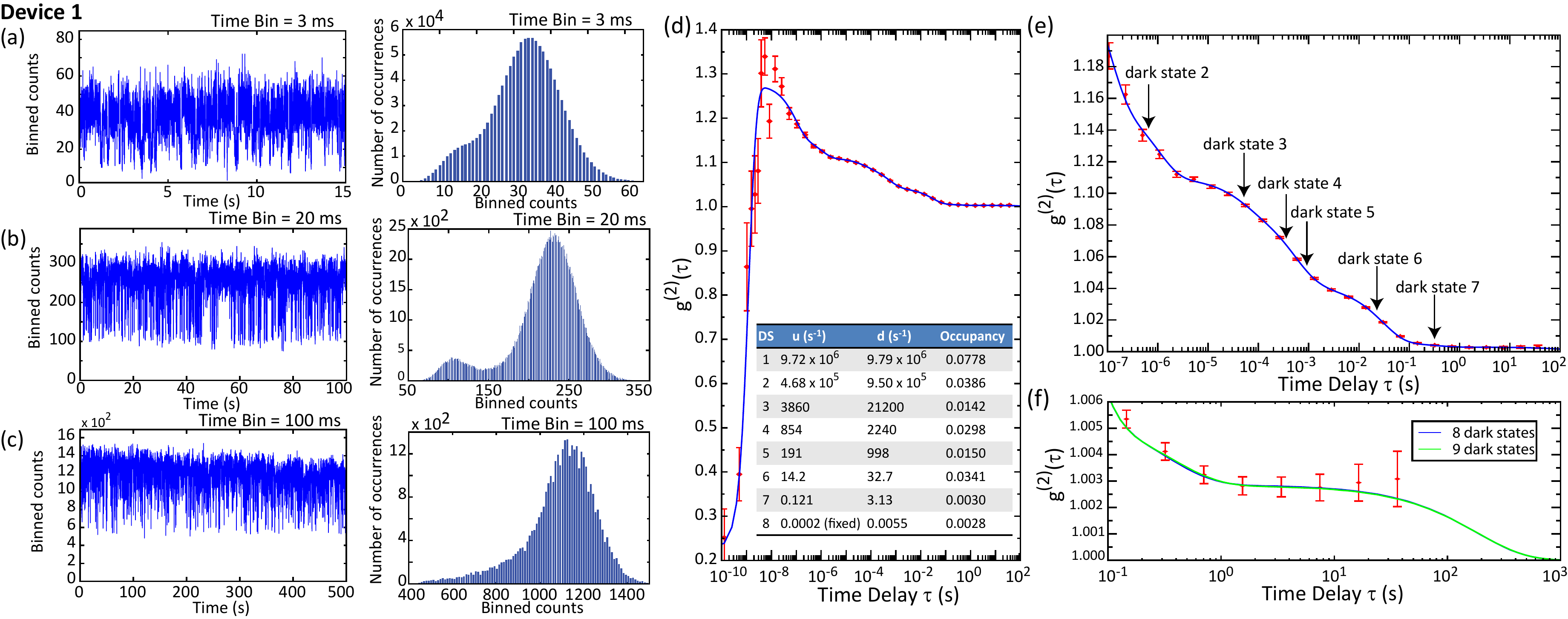}} \caption{Device 1 data.
Time-trace (left) and histogram data (right) for (a) 3~ms, (b) 20~ms, and (c)
100~ms time bins. (d) $g^{(2)}(\tau)$; red points: experimental data; blue
solid line: nonlinear least-squares fit to the $N$ dark state model in
Fig.~\ref{fig:Fig1}(b), with $N$=8.  Inset table: fit values for
excitation ($u$) and decay ($d$) rates and occupancy of each dark state (DS).
(e) Same as (d), but zoomed-in to the region between $\tau$=100~ns and
$\tau$=100~s.  Times $\tau_{i}=1/(u_{i}+d_{i})$ are used as labels for the dark
states, approximately indicating the locations of maximal slope in a plot of
$g^{(2)}(\tau)$ vs. $\text{log}(\tau)$. (f) Comparison of model
for $N$=8 and $N$=9, in the region between $\tau$=10~ms and
$\tau$=1000~s. Within the region for which experimental data is available, no
significant difference is seen for $N>8$. The
estimated QD radiative efficiency is $78~\%$.} \label{fig:Fig2}
\end{figure*}

Our samples consist of InAs QDs embedded in the center of a 190~nm thick GaAs
layer.  The collection efficiency of emitted photons is enhanced through the
use of a circular grating microcavity\cite{ref:Davanco_BE,ref:Ates_JSTQE} or
fiber-coupled microdisk cavity geometry\cite{ref:Ates_Srinivasan_SciRep}, as
detailed in the Supplementary Material\cite{ref:QD_blinking_note}. The
devices are cooled to 10~K in a liquid helium flow cryostat and excited by a 780~nm (above the GaAs bandgap) continuous wave laser.
The collected emission is spectrally filtered (bandwidth $<0.2$~nm
$\approx250$~$\mu$eV) to select a single state of a single QD (typically
the bi-exciton or neutral exciton state), split
on a 50/50 beamsplitter, and sent to a pair of silicon single-photon counting
avalanche diodes (SPADs) (see Fig.~\ref{fig:Fig1}(a)). The SPAD outputs are directed to a time-correlator
that records photon arrival times for each channel with a resolution of 4~ps.
Data is typically recorded over a period of 1~h. Results from three different
devices (labeled 1,2,3) are described below.

Figure~\ref{fig:Fig2}(a) shows a portion of the fluorescence time trace
recorded from device 1, where detection events are binned into 3~ms bins. The time
trace data clearly shows a fluctuating fluorescence intensity, more often exhibiting high than low levels.
This is further seen in the histogram of detection events per bin shown next to the time
trace, which exhibits a bimodal behavior biased towards the higher count
rates. Figures~\ref{fig:Fig2}(b) and ~\ref{fig:Fig2}(c) show the
corresponding time trace and histogram data for 20~ms and 100~ms time bins.
At the 20~ms bin width, a bimodal distribution in the histogram is more clearly seen.
At 100~ms, it becomes less visible, as is the contrast between high and low
intensities in the time trace.  When the bin width is increased to
1000~ms, obvious signs of blinking are no longer observable
in either the time trace or histogram data\cite{ref:QD_blinking_note}.

The sensitivity of time trace and histogram data to the choice of bin width
is
well-known\cite{ref:Lippitz_Orrit_blinking_review,ref:Verberk_Orrit_JCP,ref:Crouch_Pelton_blinking},
and can limit the ability to achieve a complete picture of the system
dynamics.  In particular, the minimum reliable bin size is limited by the
available photon flux and the shot noise, while too large bin sizes average
out fluctuations occurring at shorter time scales. Histogram analysis of the
high or low fluorescence level time interval distributions is furthermore
influenced by the selection of a threshold intensity level.  In contrast,
intensity autocorrelation analysis does not require such
potentially arbitrary input parameters.  In Fig.~\ref{fig:Fig2}(d), we plot
the intensity autocorrelation function $g^{(2)}(\tau)$ for device 1 over a
time range exceeding eleven orders of magnitude (error bars are due to fluctuations in the detected photon count
rates, and represent one standard deviation\cite{ref:QD_blinking_note}).  This data, calculated
using an efficient approach\cite{ref:QD_blinking_note} similar to that
described in ref.~\onlinecite{ref:Laurence_photon_correlation}, indicates
that the photon antibunching at $\tau$=0, expected for a single-photon
emitter, is followed by photon bunching that peaks in the 10~ns
region before slowly decaying, with $g^{(2)}(\tau)=1$ only occurring for
$\tau>0.1$~s (see zoomed-in data in Fig.~\ref{fig:Fig2}(e)). The decay in
$g^{(2)}(\tau)$ is punctuated by a series of inflection points ('shoulders')
in which the concavity of the curve changes. Such features have been observed
in fluorescence autocorrelation curves of single aromatic molecules in
polymeric
hosts\cite{ref:zumbush,ref:Fleury_Orrit_spectral_diffusion_single_molecule}.
Photon bunching in these systems arises from shelving of the molecule into
dark triplet states, resulting in bursts of emitted photons followed by dark
intervals at characteristic rates.  Similar behavior can also originate from
interactions between the molecule and neighboring two-level systems (TLS) in
the host polymer. Switching between the states of the TLS leads to sudden
jumps in emission frequency, and correspondingly, emission intensity.
Multiple shoulders in the autocorrelation have been associated with coupling
to a number of TLSs with varying switching rates.

We interpret our $g^{(2)}(\tau)$ data similarly, taking the
radiative QD transition to be coupled to multiple non-radiative, or dark,
states\cite{ref:QD_blinking_note}, as depicted in Fig.~\ref{fig:Fig1}(b).
This phenomenological model is motivated by potential physical mechanisms
present in self-assembled InAs/GaAs QDs.   For example, lattice defects in
the vicinity of the QD can act as carrier traps, and charge tunneling
events between the QD and such traps lead to fluorescence
intermittency\cite{ref:Wang_Shih_blinking,ref:Pistol_Samuelson_PRB}.
Perturbation of the electron and hole wavefunction overlap by the local
electric field of trapped charges has also been postulated as a cause of
blinking\cite{ref:Sugisaki}. Another possibility is that tunneling of
carriers into nearby traps causes spectral shifts of the QD emission out of
the $\approx0.2$~nm filter
bandwidth, leading to an effective blinking behavior. Such
shifts would however be larger than the spectral diffusion measurements
recently reported\cite{ref:Berthelot,ref:Abbarchi_QD_spect_diffusion,ref:Vamivakas_Atature_PRL,ref:Houel_Warburton_PRL}.
We did not observe such spectral diffusion in spectroscopy with a 0.035~nm
resolution.

In these scenarios, interactions with surrounding traps drive the QD into
high or low emission states with well-defined rates, consistent with the
model of Fig.~\ref{fig:Fig1}(b). Here, each dark state $i$ is populated at a
rate $u_{i}$ and de-populated at a rate $d_{i}$. We solve the rate equations
to compute $g^{(2)}(\tau)$ using the appropriate transition for each device:
The XX $\rightarrow$ X$_{\rm B}$ transition for devices 1 and 2, and the
X$_{\rm B}$ $\rightarrow$ G transition for device 3. All parameters are
varied in the fit except for the radiative decay rate, $r_{\text{down}}$,
which is determined from independent
measurements\cite{ref:QD_blinking_note}. A first estimate of the number of
dark states used in the model is the number of shoulders in the measured
$g^{(2)}(\tau)$ data.  Ultimately, the number of dark states is determined by
the $\chi^2$ parameter minimized in the fit, as defined in\cite{ref:QD_blinking_note}.

Fits to device 1 data are shown as blue solid lines in
Figs.~\ref{fig:Fig2}(d)-(e), along with extracted occupancy and
population and de-population rates, $u_{i}$ and $d_{i}$, of each dark state.
The short ($<10$~ns) time behavior of $g^{(2)}(\tau)$ depends primarily on
the excitation rate $r_{\rm up}$, the decay rate $r_{\rm down}$, the SPAD
timing jitter, and the background signal, if
present\cite{ref:QD_blinking_note}. The behavior at longer times depends
primarily on $u_{i}$ and $d_{i}$. Accurate fits
require a minimum number $N$ of dark states (8 for this device), below which the
behavior of $g^{(2)}(\tau)$, quantified by the fit $\chi^2$, is not well reproduced. Larger $N$ has negligible impact
on the fit (Fig.~\ref{fig:Fig2}(f)), and the total dark state occupancy
changes by $<0.05~\%$. The radiative quantum yield (radiative efficiency)
is estimated by subtracting this total dark state occupancy from unity.  We note that as the rates coupling states
G, X$_{\rm B}$, X$_{\rm D}$, and XX are more than an order of magnitude
faster than rates to the dark states, the quality of the fits and the resulting quantum
yield does not depend on whether the dark states are coupled to G (Fig.~\ref{fig:Fig1}) or X$_{\rm
B}$. Similarly, replacing the dark states with partially emissive gray
states\cite{ref:Spincelli_gray_state_PRL}, modeled with a branching ratio between dark and bright transitions,
does not significantly affect the fits or computed quantum yield\cite{ref:QD_blinking_note}.

In all, the $g^{(2)}(\tau)$ data and rate equation analysis uncover
qualitatively new information about blinking in this device in comparison to
the time trace and histogram data. First, we see that blinking occurs across
a wide variety of time scales. While blinking at sub-microsecond time scales
has been reported in epitaxially grown InAs QDs
previously\cite{ref:Santori4,ref:Volz_Imamoglu}, our measurements show that
these systems can exhibit blinking out to hundreds of milliseconds. One physical picture qualitatively consistent with this observation would be that blinking is caused by the
tunneling of carriers between the QD and several adjacent traps of varying separation
from the QD.  For example, Sercel and colleagues have considered electron relaxation
from a QD through a deep level trap~\cite{ref:Sercel,ref:Schroeter_Griffiths_Sercel},
and calculated that tunneling rates can vary by several orders of magnitude over a few
tens of nanometers of QD-trap separation (see Supplementary Material for a plot of these tunneling
rates).  Such deep level traps may arise during the QD growth process
itself~\cite{ref:Sercel,ref:Lin_Song_QD_deep_levels,ref:Asano_deep_traps_QDs} may physically
correspond to impurities, such as vacancies, antisites, and interstitials, produced during
growth and post-growth fabrication processes. We point out that a rapid thermal annealing step was
used to blue-shift the QD emission in our wafers~\cite{ref:QD_blinking_note,ref:Lin_deep_levels_QD_RTA}, prior to device fabrication.

The estimated total dark state occupancy is $21.6~\%$, so the radiative transition is still dominantly preferred over excitation into the
dark states, with a radiative quantum yield of $78.4$~$\%$.
Finally, we note that the rates $u_{i}$ and $d_{i}$ for
populating/de-populating dark state $i$ can at least be qualitatively linked
to the location of prominent features in the $g^{(2)}(\tau)$ data.  For
example, in a system consisting of a single dark state that is populated and
de-populated at rates $u_{1}$ and $d_{1}$, the slope of $g^{(2)}(\tau)$
plotted vs. $\text{log}(\tau)$ is maximimal at
$\tau=1/(u_{1}+d_{1})$\cite{ref:Lippitz_Orrit_blinking_review}. In a system
comprised of multiple dark states, if excitation and decay rates are
sufficiently different, the values $\tau_{i}=1/(u_{i}+d_{i})$ still
approximately point to slope maxima.  Figure~\ref{fig:Fig2}(e)
identifies these points, which qualitatively match the experimentally
observed maximum slope points. Quantitative details are given in\cite{ref:QD_blinking_note}.

\begin{figure}[t]
\centerline{\includegraphics[width=8.5 cm]{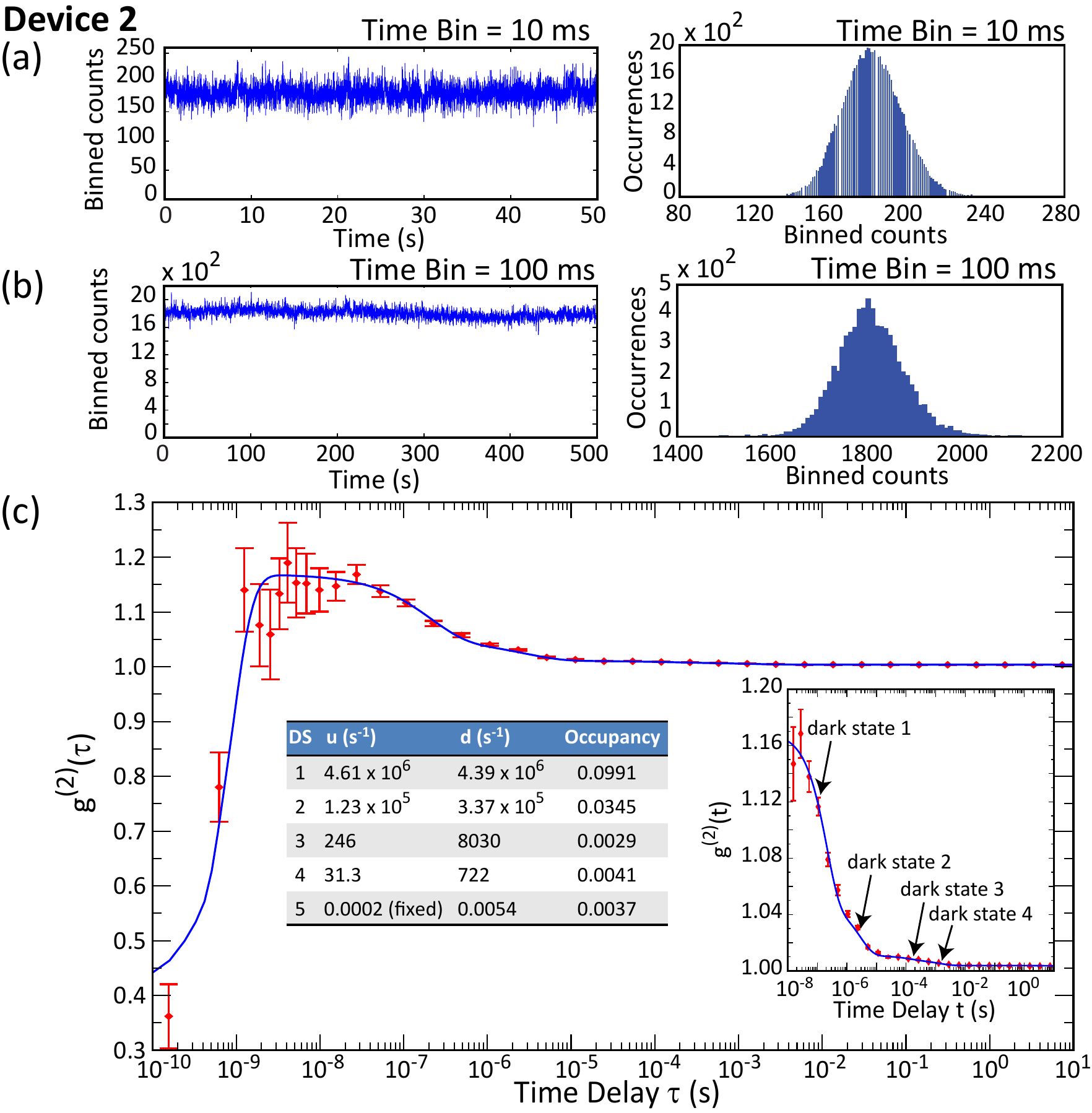}}
\caption{Device 2 data.  Time-trace (left) and histogram data (right) for (a) 10~ms and
(b) 100~ms time bins. (c) $g^{(2)}(\tau)$; red points:
experimental data; blue solid line: nonlinear least-squares fit to the
model with $N$=5. Inset table: fit values for excitation ($u$) and decay ($d$) rates, and occupancy of each dark state (DS). Inset graph: data over $\tau$=[10~ns, 10~s]. Points $\tau_{i}=1/(u_{i}+d_{i})$ are indicated for each dark state.
The estimated QD radiative efficiency is $86~\%$.}
\label{fig:Fig3}
\end{figure}

Repeating this analysis for device 2 yields the results in
Fig.~\ref{fig:Fig3}.  Here, neither time trace nor histogram data show
clear evidence of blinking.  In contrast, $g^{(2)}(\tau)$ in
Fig.~\ref{fig:Fig3}(c) again evidences bunching at the ten nanosecond time
scale, followed by a series of shoulders, before reaching a value of unity.
The data is fit to a rate equation model with five dark states, and
again shows close correspondence (inset to Fig.~\ref{fig:Fig3}(c)). Contrasting with device 1,
blinking at longer times (e.g.,
$>10~\mu$s) is significantly less pronounced, and the estimated total dark state
occupancy is $14.4~\%$.

\begin{figure}[t]
\centerline{\includegraphics[width=8.5 cm]{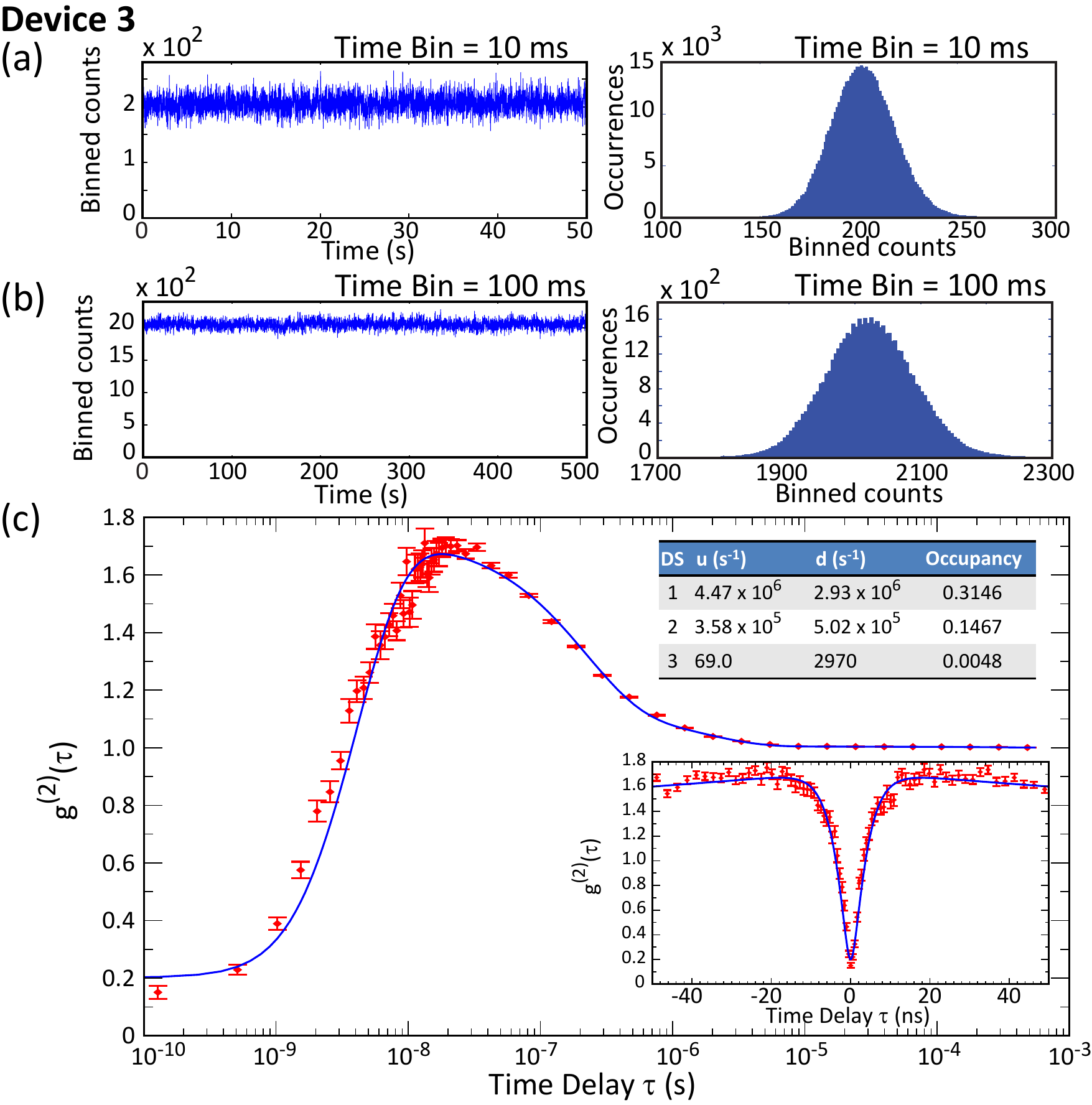}}
\caption{Device 3 data.  Time-trace (left) and histogram data (right) for (a) 10~ms and
(b) 100~ms time bins. (c) $g^{(2)}(\tau)$; red points:
experimental data; blue solid line: nonlinear least-squares fit to the
model with $N$=3.  Inset table: fit values for excitation ($u$) and
decay ($d$) rates and occupancy of each dark state (DS). Inset plot: data with linear time scale showing the anti-bunching dip at $\tau=0$; the pronounced photon bunching away from zero delay is not evident on this scale. The estimated QD radiative efficiency is $53~\%$.}
\label{fig:Fig4}
\end{figure}

Finally, we present data from device 3 in Fig.~\ref{fig:Fig4}.  Similar to
device 2, the time trace and histogram data show little evidence of blinking.
The $g^{(2)}(\tau)$ data does reveal significant
blinking over sub-microsecond time scales, but at longer times blinking is
minimal and the system can be well fit to a model with $N=3$
(with final state occupancy $<0.5~\%$).
Interestingly, the total dark state occupancy is
$46.7~\%$, significantly greater than observed in either device 1
or 2.  Thus, despite qualitative similarity with the time trace and histogram
data of device 2, the dynamics of the QD are in fact qualitatively different,
as revealed by the photon correlation measurements. This qualitative difference is perhaps
unsurprising given its entirely different device history (different wafer growth; no
rapid thermal annealing~\cite{ref:QD_blinking_note}.  Also, as
the pronounced bunching persists out to $\mu$s time scales, an
accurate estimate of $g^{(2)}(0)$, needed for assessing the purity of the single-photon source,
requires acquisition and analysis of data out to many orders of
magnitude longer times than the characteristic time scale of the antibunching
dip.

Correlation functions can reveal the kinetics of the blinking signal over a large time
range, however only in a time-averaged sense\cite{ref:Lippitz_Orrit_blinking_review}. Information about instantaneous
intensity fluctuations, such as probability distributions for bright and dark intervals, can be obtained from
photon-counting histograms, as commonly done in
the blinking literature\cite{ref:Lippitz_Orrit_blinking_review,ref:Stefani_PT_blinking,ref:Crouch_Pelton_blinking}.
Applied to epitaxially-grown QDs, this type of analysis has revealed exponential blinking time distributions\cite{ref:Pistol_Samuelson_PRB,ref:Sugisaki,ref:Wang_Shih_blinking},
suggesting modification of the QD fluorescence by one or a few neighboring centers, as discussed
(nanocrystal QDs have in contrast been
shown to display power-law distributions\cite{ref:Crouch_Pelton_blinking}).
We have applied this technique to the QD in Device 1. Although our
measured data does not strictly follow the stringent criteria suggested
in\cite{ref:Crouch_Pelton_blinking} for reliable parameter extraction, we
see strong indications of exponential probability distributions\cite{ref:QD_blinking_note}.

In summary, photon correlation measurements taken over eleven orders of
magnitude in time are used to study blinking in epitaxially-grown,
self-assembled InAs QDs housed in photonic nanocavities.  The
measurements are fitted to a rate equation model consisting of a
radiative transition coupled to a number of dark states.  The model
reproduces the observed behavior, allowing us to quantify the multiple
blinking time scales present and estimate the QD radiative
efficiency, which ranges between 53$~\%$ and 85$~\%$.  We anticipate that this approach will be valuable in studying the
behavior of InAs QDs in proximity ($<100$~nm) to etched surfaces and/or metals in
nanophotonic/nanoplasmonic geometries. Indeed, the
blinking observed here may stem from traps produced in the
fabrication of the nanostructures used to enhance QD emission collection.  Measuring photon correlations across this broad
range of time scales both before and after nanofabrication may help
elucidate the origin of blinking in these systems.

C.S.H. acknowledges the CNST Visiting Fellow program and support from the Office of Naval Research through the Naval Research Laboratory's Basic Research Program. M.D. and S.A. acknowledge support under the Cooperative Research Agreement between the University of
Maryland and NIST-CNST, Award 70NANB10H193.

\bibliographystyle{apsrev4-1}
%\bibliography{KS_bib_2013_12_17}
%

\title{Supplementary Material for \\ Multiple time scale blinking in InAs quantum dot single-photon sources}% Force line breaks with \\

\author{Marcelo Davan\c co}\email{marcelo.davanco@nist.gov}
\affiliation{Center for Nanoscale Science and Technology, National Institute
of Standards and Technology, Gaithersburg, MD 20899,
USA}\affiliation{Maryland NanoCenter, University of Maryland, College Park,
MD}
\author{C. Stephen Hellberg}\email{steve.hellberg@nrl.navy.mil}
\affiliation{Center
for Computational Materials Science, Code 6390, Naval Research Laboratory,
Washington DC 20375}
\author{Serkan Ates}
\affiliation{Center for Nanoscale Science and Technology, National Institute
of Standards and Technology, Gaithersburg, MD 20899, USA}
\affiliation{Maryland NanoCenter, University of Maryland, College Park, MD}
\author{Antonio Badolato}
\affiliation{Department of Physics and Astronomy, University of
Rochester, Rochester, New York 14627, USA}
\author{Kartik Srinivasan}\email{kartik.srinivasan@nist.gov}
\affiliation{Center for Nanoscale Science and Technology, National
Institute of Standards and Technology, Gaithersburg, MD 20899, USA}
\date{\today}% It is always \today, today,

\newpage

\section{Supplemental Information}% Force line breaks with \\

\setcounter{figure}{0}
\makeatletter
\renewcommand{\thefigure}{S\@arabic\c@figure}

\setcounter{equation}{0}
\makeatletter
\renewcommand{\theequation}{S\@arabic\c@equation}

\section{Device Details}

Devices based on two different quantum dot (QD) wafers are studied in this
work. Devices 1 and 2 come from wafer A, which is made of an epistructure
consisting of a 190~nm thick GaAs waveguide layer on top of a 1 $\mu$m thick,
Al$_{0.6}$Ga$_{0.4}$As sacrificial layer.  The as-grown emission wavelength
of the QD s-shell is near 1100~nm, so the samples are blue-shifted to a
wavelength near 940~nm using a rapid thermal annealing
process~\cite{ref:Malik} performed at 830$^{\circ}$C, so that the QD emission
can be detected using efficient Si single-photon avalanche diodes (SPADs).
Circular grating microcavities~\cite{ref:Davanco_BE,ref:Ates_JSTQE} are
fabricated in order to increase the fraction of QD emission collected by the
0.42 numerical aperture (NA) lens used in the confocal microscope setup.

\begin{figure}[t]
\centerline{\includegraphics[width=\linewidth]{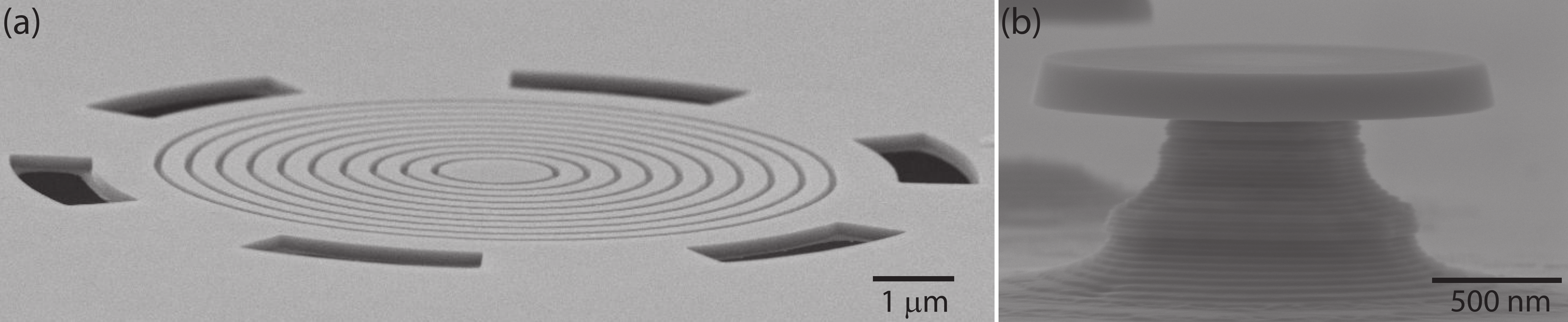}}
 \caption{Scanning electron microscope images of the (a) circular grating microcavity and (b) microdisk cavity geomtry used in this work to efficiently extract photons from a single quantum dot.}
\label{fig:SFig1}
\end{figure}

\begin{figure}[t]
\centerline{\includegraphics[width=\linewidth]{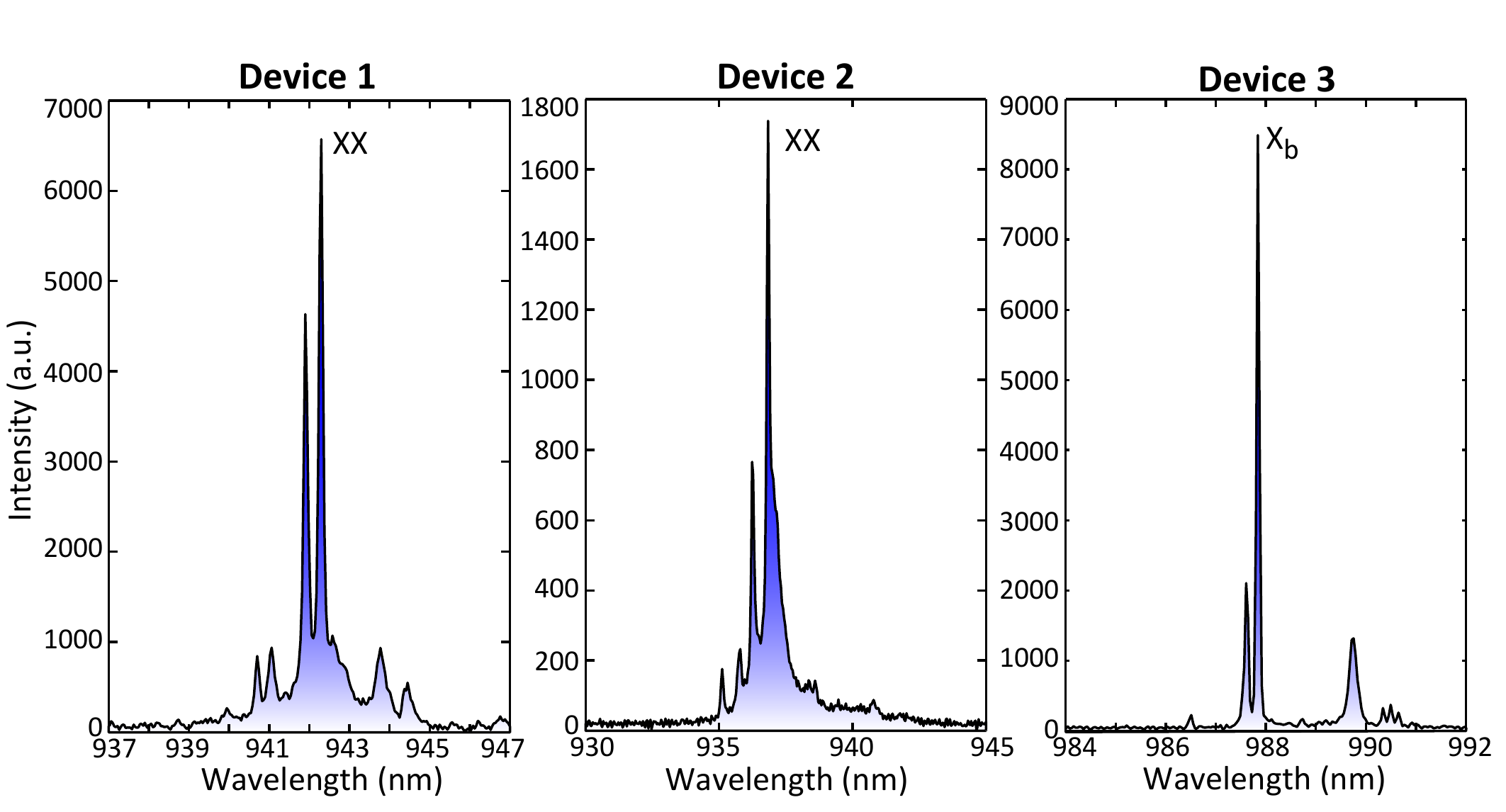}}
\caption{Photoluminescence spectra for Devices 1, 2, and 3, under 780~nm cw excitation near saturation.  The specific QD transition that is spectrally selected for photon counting measurements (XX for Devices 1 and 2, X$_{\text{b}}$ for Device 3) is indicated in each spectrum.}
\label{fig:SFig2}
\end{figure}

Device 3 comes from wafer B, which is made of a similar epistructure as wafer
A, but has an as-grown QD s-shell emission wavelength near 980~nm, so that no
rapid thermal annealing process is performed on this wafer (in comparison to Devices 1 and 2,
we note that the lack of a rapid thermal annealing step may have a significant effect
on the behavior of deep level traps in the material~\cite{ref:Lin_deep_levels_QD_RTA}). Microdisk
cavities are fabricated to increase the collection efficiency of emitted
photons, with a fiber taper waveguide interface used to extract photons from
the microdisk optical mode~\cite{ref:Ates_Srinivasan_SciRep}.  Scanning
electron microscope images of the fabricated devices are shown in
Fig.~\ref{fig:SFig1}.  Figure~\ref{fig:SFig2} shows the fluorescence spectrum of
each device, under the non-resonant 780~nm cw excitation conditions used in the photon
counting measurements presented in the main text.

Devices 1 and 2 were previously studied under pulsed (80~MHz rep rate),
820~nm non-resonant excitation in Ref.~\cite{ref:Ates_JSTQE}, where the
performance of the devices as triggered single-photon sources was assessed.
There, a photon collection efficiency of 9.5~$\%$ and 5.6~$\%$ into the
NA=0.42 lens was estimated for the two devices, respectively, under the
assumption of unity radiative quantum yield. As the radiative quantum yields
we estimate in the main text of the current work are less than unity (78~$\%$
and 86~$\%$), this indicates that the collection efficiencies determined in
the prior work were likely underestimated (by approximately 28~$\%$ and
16~$\%$, respectively).  This correction is tentatively presented, however,
both because the data for the two works were acquired on different occasions,
and more importantly, because the excitation conditions differ (different
wavelength and pulsed vs. cw pumping).  Finally, we note that while Device 3
comes from the same chip as the fiber-coupled microdisk single-photon source
studied in Ref.~\cite{ref:Ates_Srinivasan_SciRep}, it is not precisely the
same device, and had an estimated collection efficiency that was about an
order of magnitude lower (on the order of 1~$\%$ into one channel of the
optical fiber).

\section{Experimental details}

The devices were cooled to 10~K in a liquid helium flow cryostat and excited,
non-resonantly, by a 780~nm (above the GaAs bandgap) continuous wave laser.
Devices 1 and 2 were excited with a free-space beam though a NA 0.42
objective. Photoluminescence (PL) emitted by QDs contained in the circular
grating microcavities was collected through the same objective and focused
into a single-mode optical fiber, to be routed towards
detection~\cite{ref:Ates_JSTQE}. Quantum dots in Device 3 were optically
excited with a fiber taper waveguide (FTW) evanescently coupled to the
hosting microdisk, and the emitted PL was collected with the same optical
fiber~\cite{ref:Ates_Srinivasan_SciRep}. In both situations, the collected PL
was spectrally filtered to select only a single state of a single QD (the
bi-exciton or neutral exciton state - see Fig.~\ref{fig:SFig2}). For devices
1 and 2, a monochromator was used as a filter, with a bandwidth of
$\approx100$~pm. In the case of Device 3, two cascaded holographic gratings
were used, providing an overall bandwidth of $\approx200$~pm. The filtered PL
was split on a 50/50 beamsplitter and sent to a pair of silicon single-photon
counting avalanche diodes (SPADs). The SPAD outputs were directed to a
time-correlator that recorded photon arrival times for each channel with a
resolution of 4~ps. The SPAD timing jitter was $\approx500$~ps.

\section{Histogram data}

The measured data consisted of a series of time tags corresponding to single-photon detection events. To produce the time-domain blinking
trajectories shown in the main text, detection events in the raw time-tag
data were placed into a sequence of bins of duration $\Delta t$. Histograms
for number of detection events per bin were generated from the blinking
trajectories. The number of histogram bins was selected through the
Freedman-Diaconis method~\cite{ref:Freedman_Diaconis}.

\begin{figure}[t]
\centerline{\includegraphics[width=\linewidth]{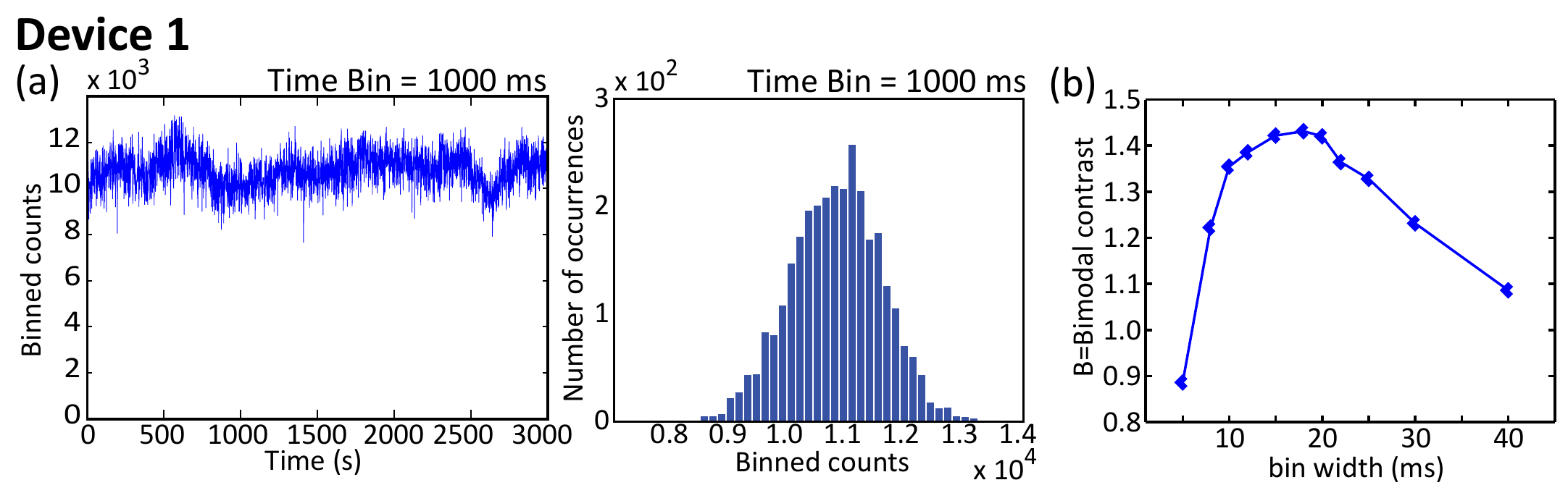}}
\caption{Device 1 supplemental
data. (a) Time-trace (left) and histogram data (right) for 1000~ms time bins,
(b) Bimodal contrast parameter as a function of time bin width.}
\label{fig:SFig3}
\end{figure}

In Fig.~2 in the main text, we show time trajectory and histogram data for
Device 1 using time bin widths of 3~ms, 20~ms, and 100~ms.  This represents
essentially the full range of bin widths over which appreciable blinking can
be directly observed in trajectory and histogram data.  For example, a bin
width of 1000~ms shows little variation in the collected emission intensity
in the time trajectory, and the histogram data does not show evidence of a
bimodal distribution (Fig.~\ref{fig:SFig3}(a)).

For time bin widths over which a bimodal distribution is visible, we define a
contrast parameter $B=|1-I_{\text{min}}/I_{1}|+|1-I_{\text{min}}/I_{2}|$,
where $I_{1}$ and $I_{2}$ are the intensities of the two maxima and
$I_{\text{min}}$ is the minimum intensity between the two peaks.  In
Fig.~\ref{fig:SFig3}(b), we plot $B$ as a function of time bin width for
Device 1.  We see that there is a relatively small range of time bin widths
over which a bimodal distribution is evident, with the contrast optimized at
$\approx20$~ms, for which data was plotted in the main text in Fig.~2(b).

\begin{figure*}[t]
\includegraphics[width=10.5 cm]{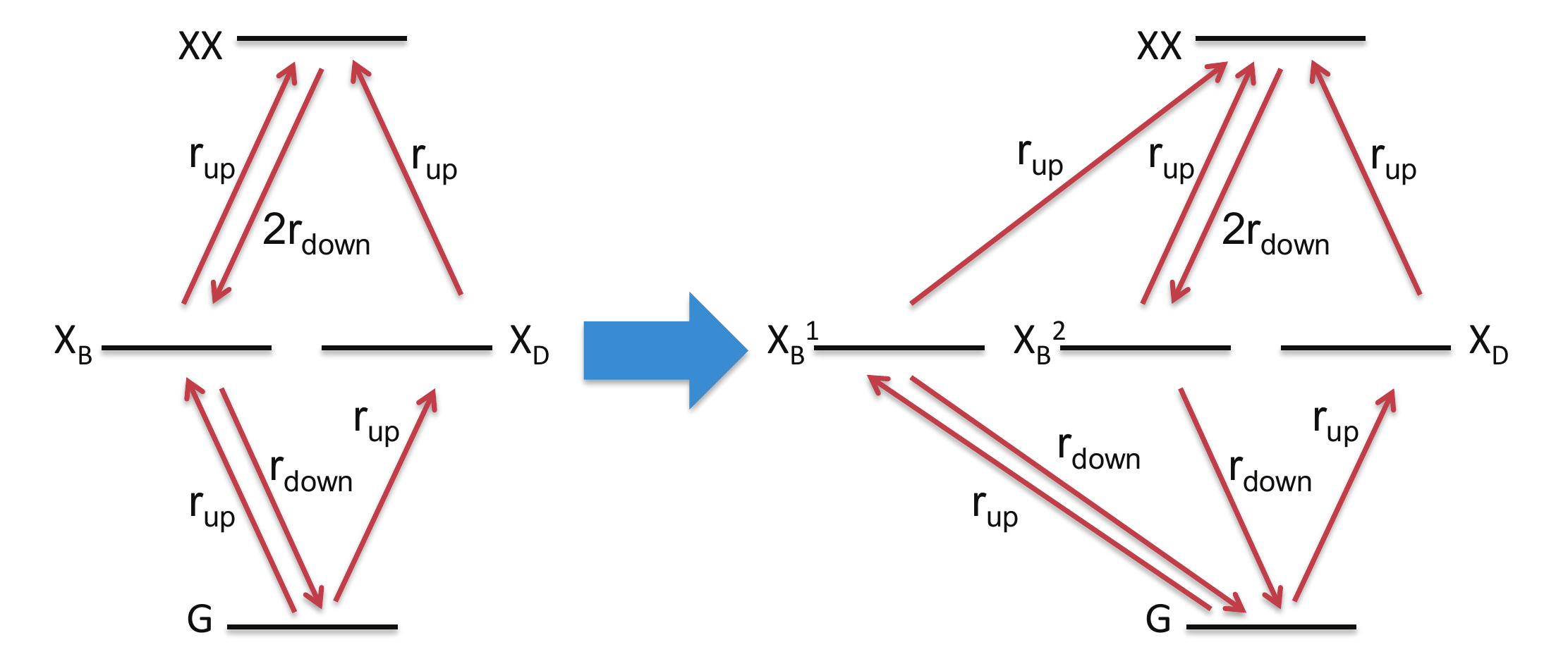}
\caption{States and transitions for the computation
of $g^{(2)}(\tau)$ for the XX $\rightarrow$ X$_{\rm B}$ transition.
The state ${\rm X_B}$ is split into two states: ${\rm X_B^1}$ and ${\rm X_B^2}$.
With
initial conditions of $p_{\rm X_B^1}(\tau=0)=1$, the correlation function is
simply given by $g^{(2)}(\tau) = p_{\rm X_B^2}(\tau)$. Dark states are not
shown in these diagrams for clarity.}\label{fig:SFig4}
\end{figure*}

\section{Calculation of $g^{(2)}(\tau)$ from time-tagged data}

The experimental setup yields two streams of data corresponding
to photon arrival events on SPADs 1 and 2, collected over a period of 1 h.
These two data streams were used to calculate the second-order
correlation function
\begin{equation}
g^{(2)}(\tau) = \frac{\langle I_1(t)I_2(t+\tau)\rangle}{\langle I_1(t)\rangle\langle I_2(t)\rangle},
\end{equation}
where $I_1(t)$ and $I_2(t)$ are the photon fluxes in SPADs 1 and 2.
To calculate $g^{(2)}(\tau)$,
we use an efficient approach similar to that in
Ref.~\onlinecite{ref:Laurence_photon_correlation}.
We initially specify $M$ non-overlapping time
windows $[\tau_1^{\rm min},\tau_1^{\rm max}), [\tau_2^{\rm min},\tau_2^{\rm
max}), \ldots [\tau_M^{\rm min},\tau_M^{\rm max})$, with $\tau_{i+1}^{\rm min} = \tau_i^{\rm max}$, which are used to bin detection events in the SPAD 1 stream. The arrival times of all the photons in each of the M time windows is initially stored. When the time $t$ is shifted to the next arrival time in the SPAD 2 stream, photon numbers in all time windows in the SPAD 1 stream are updated, shifting on average 1 photon from window $i$ to window $i+1$. Thus the
overall computation scales as $O(NM)$ for $N$ photons and $M$ time windows instead of
the $O(N^2)$ cost of a direct computation of $g^{(2)}(\tau)$.

To determine the uncertainty of the calculated $g^{(2)}(\tau)$ due to signal
fluctuations caused by varying experimental conditions, the original data is
divided into $n>M$ bins, to which the procedure above is applied. The
covariance matrix
\begin{equation}
C_{ij} = \frac{1}{n-1}
\left( \frac{1}{n} \sum_{k=1}^n (g^{(i)}_{k}-\bar{g}^{(i)})
(g^{(j)}_{k}-\bar{g}^{(j)}) \right)
\label{eq:covariance}
\end{equation}
is then calculated, where $g^{(i)}_{k}$ is the measured value of the $i$th
time window of $g^{(2)}$ in the $k$th statistical bin, and $\bar{g}^{(i)}$ is
the mean $g^{(2)}$ over the $n$ statistical bins (we drop the (2) superscript
in $g^{(2)}$ for clarity). The error bars shown in the plots in the main text
are the standard deviations of the individual observations: $\sigma_i =
\sqrt{C_{ii}}$.

\section{Rate equation model and fits}

The dynamics of the quantum dots are modeled by the transitions shown
schematically in Fig.~1b of the main text. The states are defined as:
\\\hspace*{1in}
\begin{tabular}{ll}
${\rm G}$ & Ground state
\\${\rm X_B}$ & Bright exciton
\\${\rm X_D}$ & Dark exciton
\\${\rm XX}$ ~ & Bi-exciton
\\${i}$ & The $i$-th dark state
\end{tabular}
\\
The populations $p$ evolve according to the rate equations:
\begin{eqnarray}
\frac{dp_{\rm XX}}{d\tau} &=&
r_{\rm up} (p_{\rm X_B}+p_{\rm X_D})-2r_{\rm down} p_{\rm XX}\nonumber
\\
\frac{dp_{\rm X_B}}{d\tau} &=&
2r_{\rm down} p_{\rm XX} + r_{\rm up} p_{\rm G}-(r_{\rm up}+r_{\rm down}) p_{\rm X_B}\nonumber
\\
\frac{dp_{\rm X_D}}{d\tau} &=&
r_{\rm up} p_{\rm G}-r_{\rm up} p_{\rm X_D}\label{rate}
\\
\frac{dp_{\rm G}}{d\tau} &=&
r_{\rm down} p_{\rm X_B} + \sum_i d_i p_i-(2r_{\rm up}+\sum_i u_i) p_{\rm X_G}\nonumber
\\
\frac{dp_{i}}{d\tau} &=& u_i p_{\rm X_G}-d_i p_i\nonumber
\end{eqnarray}

\noindent where $u_{i}$ and $d_{i}$ are the up-transition and down-transition
rates between the ground state G and the $i$th dark state.
%We note that
The state $X_D$ in the model corresponds to an electron-hole pair with parallel spins.
This state, normally called the dark exciton, is
%ideally
approximately degenerate with the bright exciton state $X_B$, however is not optically active.
The additional $N$ dark states can either be non-emissive or may emit outside the filtered detection window, as detailed in the experimental setup description.

By monitoring the power dependence of the luminescence, we determined that
the XX $\rightarrow$ X$_{\rm B}$ transition was being measured in Devices 1
and 2, while the X$_{\rm B}$ $\rightarrow$ G transition was measured in
Device 3. For Devices 1 and 2, the correlation function $g^{(2)}(\tau)$ is the
probability that two photons from the XX $\rightarrow$ X$_{\rm B}$ transition are
emitted a time interval $\tau$ apart.
To compute $g^{(2)}(\tau)$, we integrate the rate equations (\ref{rate}) setting the initial conditions to be
%In other words, by setting
$p_{\rm X_B}(\tau=0)=1$ with all other populations
zero;
% as an initial condition,
$g^{(2)}(\tau)$ corresponds to the probability that the system will return to state X$_{\rm B}$
through the XX $\rightarrow$ X$_{\rm B}$ transition~\cite{ref:carmichael76}.
To compute this probability, we split the state X$_{\rm B}$ into two states in the rate
equations as shown in Fig.~\ref{fig:SFig4}. X$_{\rm B}^2$ can only be reached
via the XX $\rightarrow$ X$_{\rm B}$ transition. All other transitions to
state X$_{\rm B}$ are directed to X$_{\rm B}^1$. All transitions away from
state X$_{\rm B}$ are included for both X$_{\rm B}^1$ and X$_{\rm B}^2$.
Initially we set $p_{\rm X_B^1}(\tau=0)=1$ with all other populations zero. Then $g^{(2)}(\tau) = p_{\rm
X_B^2}(\tau)$.

For Device 3, $g^{(2)}(\tau)$ is the
probability that two photons from the X$_{\rm B}$ $\rightarrow$ G transition are
emitted a time interval $\tau$ apart.
%Since G is only accessible via downward transitions from X$_{\rm B}$, we set $p_{\rm G}(\tau=0)=1$ with all other populations
%zero as an initial condition and find $g^{(2)}(\tau) = p_{\rm X_G}(\tau)$.
Similarly to the case above, we split state G into two states: G$^2$ which can only be reached via the
X$_{\rm B}$ $\rightarrow$ G$^2$ transition, and G$^1$, which is reached by all other transitions to G, namely
the recoveries from the dark states in Fig. 1(b).
Setting $p_{{\rm G}^1} (\tau=0)=1$ with all other populations zero, $g^{(2)}(\tau) = p_{{\rm G}^2} (\tau)$.

We find the short-time behavior of $g^{(2)}(\tau)$ can be used to determine
the ratio $r_{\rm up}/r_{\rm down}$ but not the individual values of $r_{\rm
up}$ and $r_{\rm down}$. We therefore determine the values of $r_{\rm down}$
using independent decay-rate measurements. The values of $r_{\rm up}$ and
$r_{\rm down}$ for the three devices are given in Table~\ref{r_up_r_down}.

\begin{table}
\caption{\label{r_up_r_down}Additional parameters for the three devices not
included in the main text.}
\begin{ruledtabular}
\begin{tabular}{lll}
%\hline
Device  & $r_{\rm up}(s^{-1})$  & $r_{\rm down}(s^{-1})$
\\
\hline
1 & $1.11 \times 10^9$ & $0.60 \times 10^9$
\\
2 & $2.11 \times 10^9$ & $1.25 \times 10^9$
\\
3 & $0.16 \times 10^9$ & $0.75 \times 10^9$
%\\
%\hline
\end{tabular}
\end{ruledtabular}
\end{table}

The remaining parameters were determined by a $\chi^2$ fit of $g^{(2)}(\tau)$
computed from the rate equations. For uncorrelated output, the covariance matrix ${\rm
C}$ is diagonal. With correlations, $\chi^2$ is defined as:

\begin{equation}
\chi^2 = \sum_{ij}(\bar{g}^{(i)} - g^{(i)}_{\rm fit}) {\rm C}^{-1}_{ij}
(\bar{g}^{(j)} - g^{(j)}_{\rm fit}) ,
\label{chi}
\end{equation}
where $g^{(i)}_{\rm fit}$ is the fitting function determined by solving the
rate equations (\ref{rate})~\cite{ref:hellberg00gfmc}.

%When ${\rm C}$ is diagonal, its inverse is trivial.
%For more general ${\rm C}$, small errors in its components can result in large
%errors in its inverse, so it is important to calculate ${\rm C}$ accurately.
%Increasing the number of bins decreases the statistical error in ${\rm C}$
%but increases the systematic error due to autocorrelations.
%To balance these two sources of error, we choose the number of measurements
%in each bin to be $n = M p_{\rm max}$, where $p_{\rm max}$ is the maximum
%power of the Hamiltonian.\cite{toussaint89}

\begin{figure*}[!ht]
\centerline{\includegraphics[width=\linewidth]{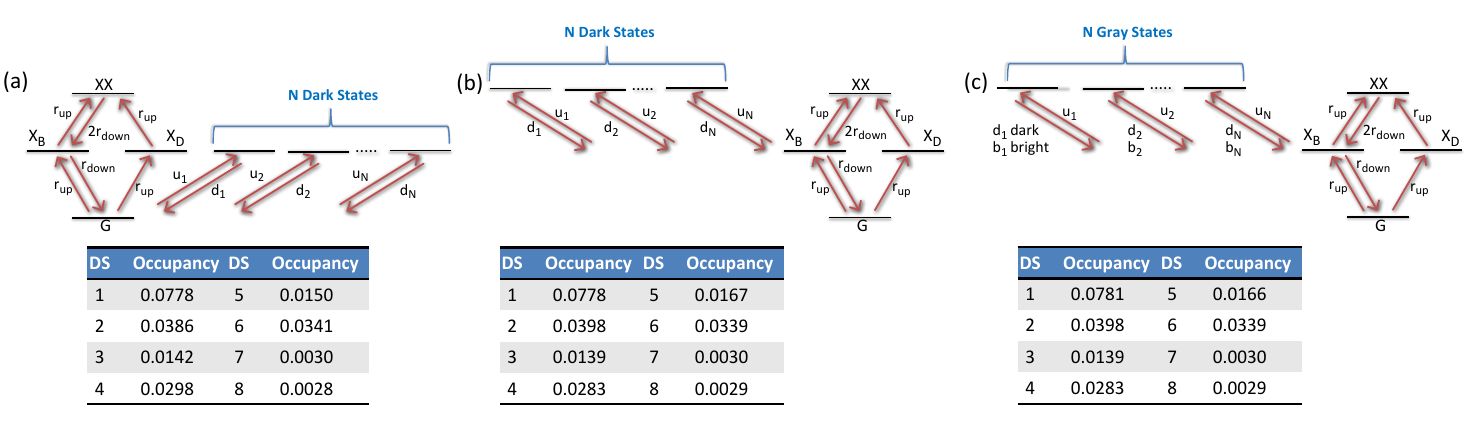}}
 \caption{Level diagrams (top) and extracted occupancies from fits (bottom) for a model with (a) dark states
 coupled to the ground state G; (b) dark states coupled to the excited state X$_{\rm B}$; (c) partially
 emissive 'gray' states emitting photons in the allowed spectral window 50~$\%$ of the time ($\alpha=1$).}
\label{fig:SFig5}
\end{figure*}

The fits found a background signal of 9~\% for Device 3; zero background was
found for Devices 1 and 2~\cite{ref:molski00}. A timing jitter of $\sigma =$
450~ps was used for all devices.

\section{Coupling of the dark states to the radiative transition}

The model employed throughout this paper has the dark states coupled to the
ground state G of the radiative transition (Fig.~1(b) in the main text).  One
might instead envision scenarios in which the dark states are coupled to
other states within the radiative transition.  We find that the fits are
relatively insensitive to which specific state within the radiative
transition they are coupled to. Physically, this is because the rates
coupling the states G, X$_{\rm B}$, X$_{\rm D}$, and XX are more than an
order of magnitude faster than rates to the dark states.

As an example, we compare the dark state occupancies determined based on fits
to the model presented in the main text (and re-displayed in
Fig.~\ref{fig:SFig5}(a)) with occupancies based on fits to a model in which
the dark states are coupled to state X$_{\rm B}$ (Fig.~\ref{fig:SFig5}(b)).
Optimizing the fits based on the $\chi^2$ parameter, we find that the
individual dark state occupancies and the total occupancy of the dark states
(and hence, the radiative efficiency of the quantum dot) are only slightly
different in the two cases.

We also considered the possibility of partially emissive 'gray' states instead
of dark states~\cite{ref:Spincelli_gray_state_PRL}.  In this model, dark state $i$
decays radiatively with rate $b_i$ (into the same spectral channel as the QD excitonic
transition) and non-radiatively with rate $d_i$, as shown schematically
in Fig.~\ref{fig:SFig5}(c).  We set the radiative fraction of each dark state to be equal,
so $b_i=\alpha d_i$.  Even with a radiative fraction as large as $\alpha=1$, the quality of the
fit and the occupancies of each dark (shown in Fig.~\ref{fig:SFig5}(c)) are nearly identical to the case with
non-emitting dark states, again a consequence of the vast difference in the rates that couple the excitonic
transition relative to the rates that couple to the dark states.

\section{Labeling of dark states}

\begin{figure}[h!]
\begin{center}
\includegraphics[width=\linewidth]{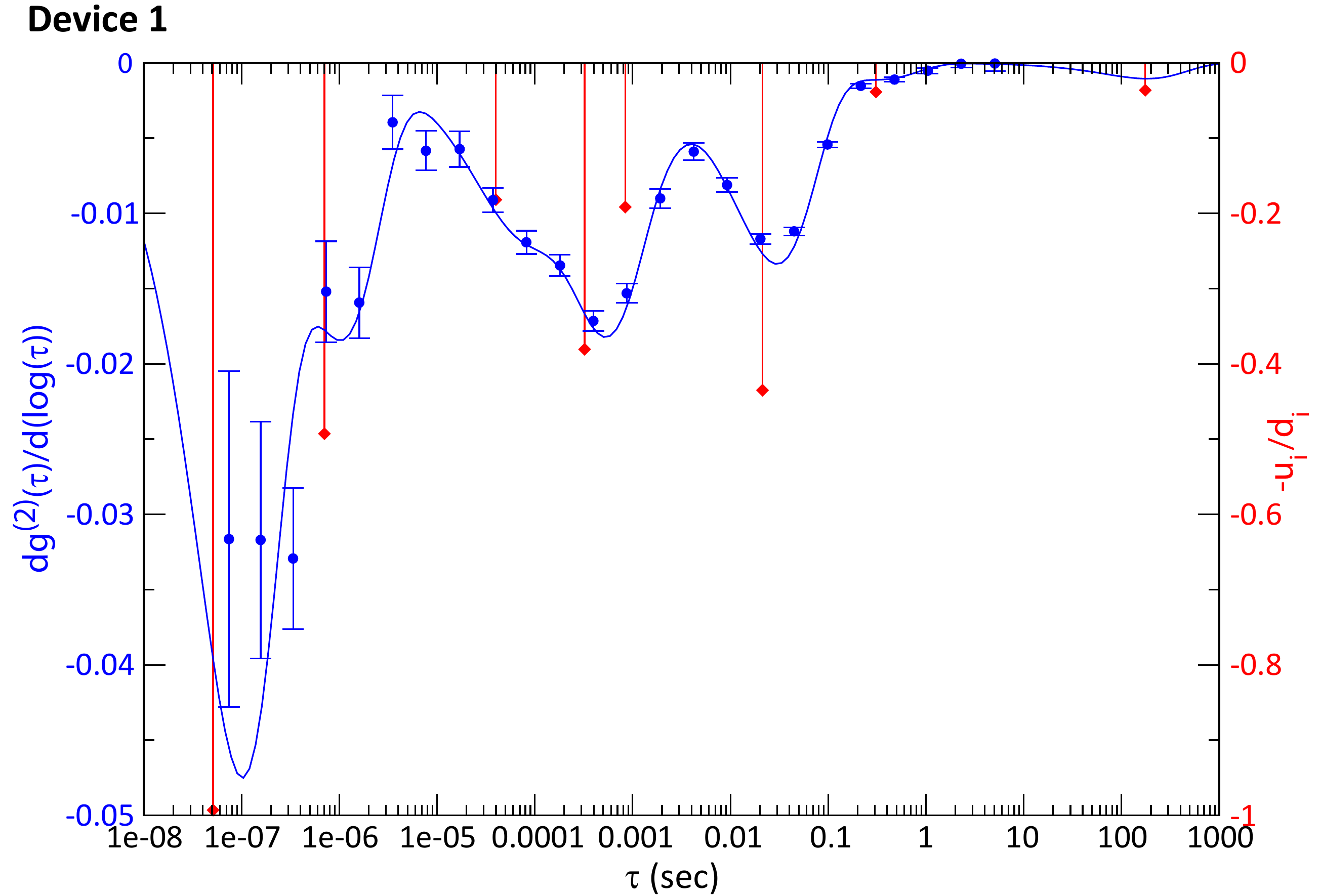}
\caption{Logarithmic derivative of $g^{(2)}(t)$, $\frac{d(g^{(2)}(\tau))}{d(\log(\tau))}$, for Device 1.  Blue points are from the measured data, while the solid blue curve is from the rate equation model fit.  The red points indicate the times $\tau_{i}=1/(u_{i}+d_{i})$, with the y-axis values given by $-u_{i}$/$d_{i}$, which is directly related to the occupancy of the dark state.}
\label{fig:SFig6}
\end{center}
\end{figure}

In the main text, we have labeled dark states in the $g^{(2)}(\tau)$ data
according to times $\tau_{i}=1/(u_{i}+d_{i})$, under the assumption that
these times correspond to points of maximal slope.  This assignment is
expected to be only approximate, since it essentially assumes that the
different dark states have no influence on each other.  We compare the
$\tau_{i}$ labeled in the above manner with the exact values at which
$g^{(2)}(\tau)$ shows maximal slope in Fig.~\ref{fig:SFig6}, and the expected
approximate correspondence is observed.  In addition, for each dark state we
also show the ratio $-u_{i}/d_{i}$, which is directly related to the
occupancy of that state.  This provides a graphical representation of which
dark states most influence the radiative efficiency of the quantum dot.

\section{Tunneling to traps}

Recent studies have suggested that deep traps form close to InAs/GaAs QDs during growth~\cite{ref:Lin_Song_QD_deep_levels,ref:Asano_deep_traps_QDs}. For example, Asano and colleagues have
studied trap densities in different regions of QD-containing material and have found their density to
increase in the presence of the QDs~\cite{ref:Asano_deep_traps_QDs}. Tunneling of carriers to these traps could cause the observed blinking.

As a plausibility check, we consider the work of Schroeter \emph{et al.}, who computed the tunneling of electrons in a spherical 5 nm radius In$_{0.5}$Ga$_{0.5}$As dot to a trap in the GaAs matrix, finding transition rates that vary enormously as the separation between the QD and trap is changed from 10 nm to 20 nm (see Fig.~3 in Ref.~\onlinecite{ref:Schroeter_Griffiths_Sercel}).

\begin{figure}[h]
\includegraphics[width=\linewidth]{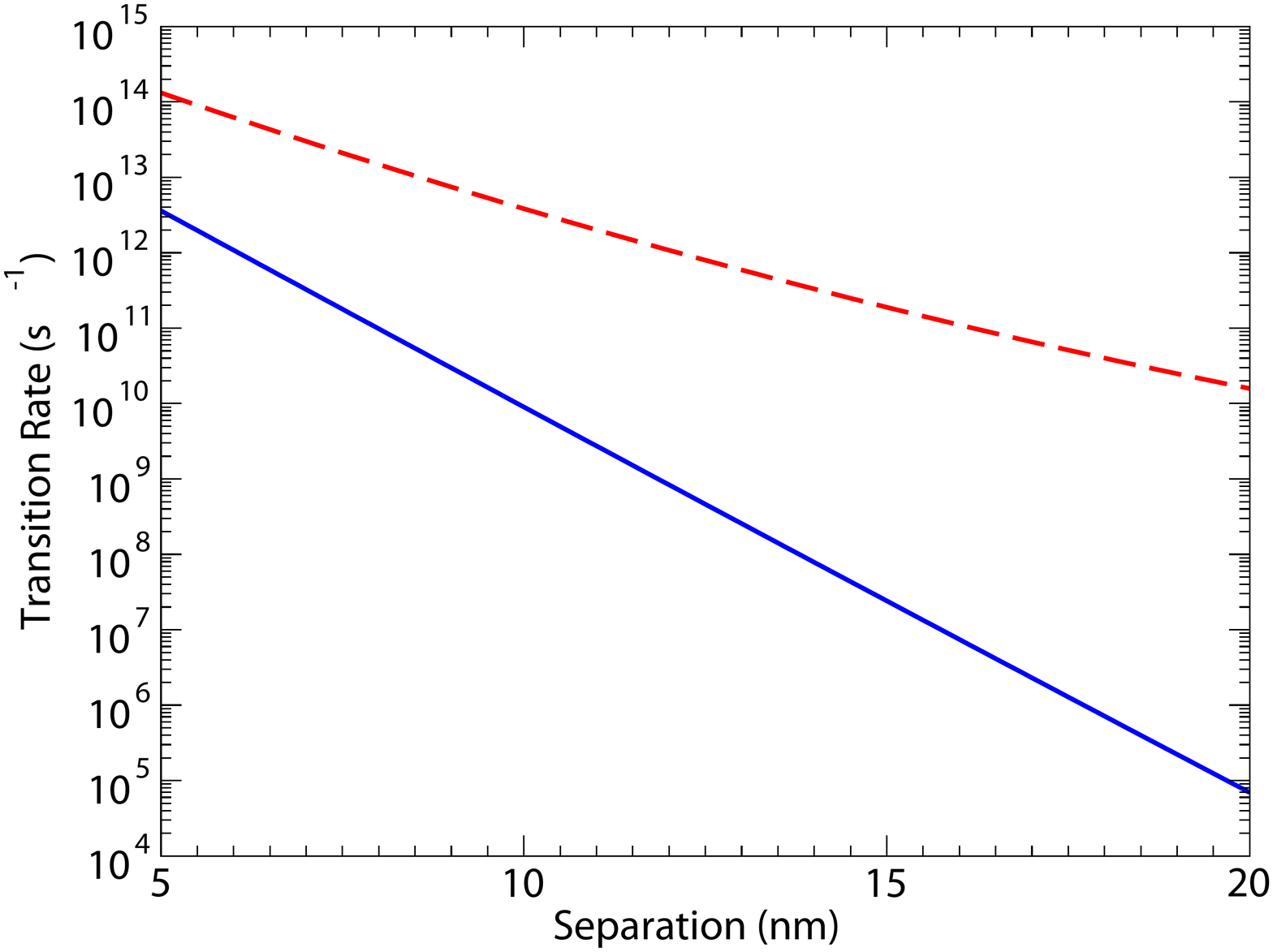}
\caption{\label{transition_rate}
Calculated transition rates between an In$_{0.5}$Ga$_{0.5}$As/GaAs QD and a nearby trap, as a function of QD-trap separation, following Ref.~\onlinecite{ref:Schroeter_Griffiths_Sercel}.  The QD excited state ($C1$) to trap ($T$) rate is shown as a red dashed line, while the the trap to QD ground state ($C0$) rate is shown as a blue solid line.}
\end{figure}

Using the parameters in Ref.~\onlinecite{ref:Schroeter_Griffiths_Sercel}, the energy level of the trap, which is modeled as a delta function potential,
lies between the ground and first excited levels of the QD. We computed the electron transition rates from the first excited
state of the QD (labeled $C1$) to the trap (labeled $T$), and from the trap to the ground state of the QD (labeled $C0$. The results as a function of the separation between the QD and the trap are shown in Fig.~\ref{transition_rate}. The $C1 \rightarrow T$ transition rate varies from $10^{14}$ s$^{-1}$ to $10^{11}$ s$^{-1}$ for dot-trap separations of 5 nm to 20 nm, while the $T \rightarrow C0$ rate varies from over $10^{12}$ s$^{-1}$ to less than $10^{5}$ s$^{-1}$.  The strong variation in tunneling rates over this distance range is consistent with the wide range of transition rates produced in the rate equation model fits to the $g^{(2)}(\tau)$ data from the main text.

Stark shifts of the QD energy levels, created when carriers populate nearby traps, could be another physical mechanism involved.  One resulting phenomenon would be spectral diffusion, that is, a fluctuation in the QD emission frequency.  For such spectral diffusion to result in an intensity fluctuation (blinking), its magnitude would need to be on the order of the optical filter bandwidth we use (250~$\mu$eV) when spectrally isolating the QD emission.  Recently, Houel and colleauges~\cite{ref:Houel_Warburton_PRL} have experimentally and theoretically studied spectral diffusion in InAs/GaAs QDs, and predict a Stark shift of $\approx$20~$\mu$eV when a single carrier (hole) is situated within a couple of nm of the QD.  Presumably multiple defect states would need to be simultaneously filled to reach shifts large enough to move the QD emission frequency outside of our filter window.  In such a scenario, a wide range of timescales for spectral diffusion could lead to the observed results.  For example, Abbarchi et al. have recently studied spectral diffusion in GaAs QDs through a correlation function approach, and fit their data to a function whose long-time scale behavior (i.e., at times much greater than the QD radiative lifetime) is given by a standard diffusion expression, meant to represent the displacement of a charge carrier near the QD~\cite{ref:Abbarchi_QD_spect_diffusion}. Plotted on a logarithmic time scale, this function shows a shoulder reminiscent of those seen in Figs.~2-4 in the main text. One could then envision a model consisting of multiple charge carriers, each causing Stark shifts to the QD emission energy and exhibiting diffusive motion with a different characteristic timescale, as producing qualitatively similar curves as the data we have measured in Figs.~2-4 in the main text.  On the other hand, the precise time-dependent behavior of the diffusive motion does not agree with our data (diffusive motion has wider tails), and the requirement of multiple diffusive timescales disagrees with the observation made in Ref.~\onlinecite{ref:Abbarchi_QD_spect_diffusion}, where a single diffusion time constant was used to model data from multiple quantum dots, and from this it was inferred that the behavior of the carrier diffusion was independent of the specific local environment.

Finally, we note that strong perturbation of the electron and hole wavefunctions due to the nearby charges has also been considered as a mechanism for blinking, by Sugisaki and colleagues~\cite{ref:Sugisaki}, and if incorporated alongside a model for trap filling dynamics, might provide an alternative explanation for what we have observed.

\section{Blinking probability distributions}

As pointed out in~\cite{ref:Lippitz_Orrit_blinking_review}, correlation
functions can reveal the kinetics of the blinking signal over a large time
range, however, they only provide averaged information about the distributions of
bright and dark time intervals. Alternatively, probability distributions for
the instantaneous intensity fluctuations can in principle be obtained from
photon-counting histograms, providing complementary information for the
characterization of single emitters. This technique is commonly done in the
blinking
literature~\cite{ref:Lippitz_Orrit_blinking_review,ref:Stefani_PT_blinking,ref:Crouch_Pelton_blinking},
and has been applied towards epitaxially-grown quantum
dots~\cite{ref:Pistol_Samuelson_PRB,ref:Sugisaki,ref:Wang_Shih_blinking}. In
the latter case, exponential blinking time distributions have been reported,
which suggests coupling of the quantum dots to neighboring two-level
systems~\cite{ref:Fleury_Orrit_spectral_diffusion_single_molecule}. Such
behavior contrasts with that of nanocrystal quantum dots, which are known to
display power-law type distributions~\cite{ref:Crouch_Pelton_blinking}.

\begin{figure}[b!]
\centerline{\includegraphics[width=8.0 cm]{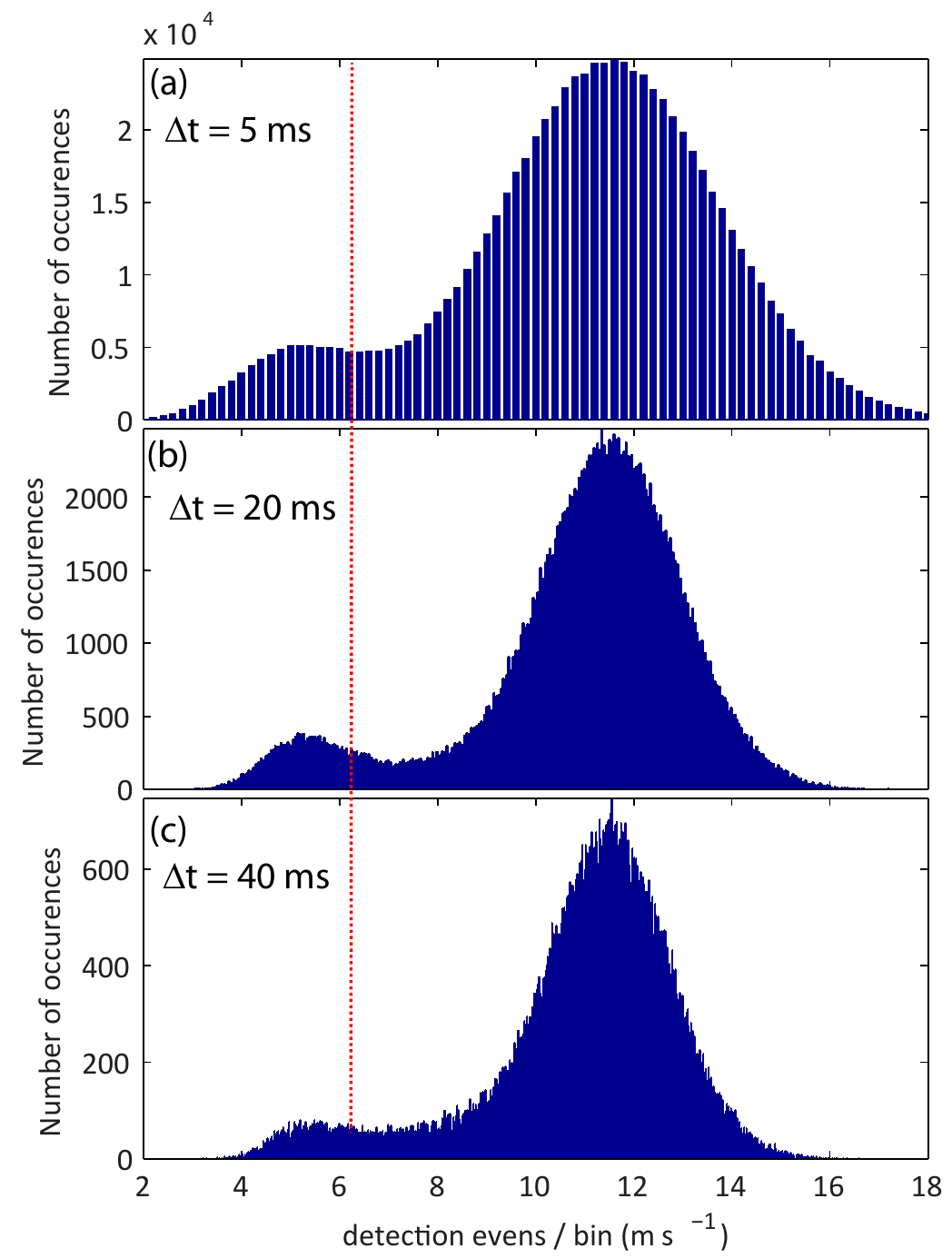}}
 \caption{Histograms of single-photon detection events per bin, for bin sizes $\Delta t$ of (a) 5 ms, (b) 20 ms and (c) 40 ms. The red dashed line indicates the threshold level used to produce the probability distributions shown in
Fig.~\ref{fig:SFig9}(a).}
\label{fig:SFig8}
\end{figure}

In what follows, we attempt to extract the probability distributions of
blinking intervals for a single quantum dot in one of our fabricated devices. The
results of Ref.~\cite{ref:Crouch_Pelton_blinking} suggest that reliable
recovery of blinking probability distribution functions require that
intensity histograms display clear bimodal distributions, with two
non-overlapping peaks corresponding to dark and bright emitter states, and
time trajectories with several thousands of blinking events; these criteria
must furthermore be met for bin sizes $\Delta t$ varying over at least a
decade. Only one of our measured devices (Device 1) produced histograms with
clear bimodal distributions, as shown in Fig.~\ref{fig:SFig8}(a)-(c), for bin
widths $\Delta t$ varying over almost a decade (from 5~ms up to 40~ms).

Although it is clear that there is some overlap between the bright and dark
histogram peaks, we proceed to analyze the probability distributions of the
'on' and 'off' time intervals of this device. We start by defining an
emission rate threshold for the binned data, below which the dot was
considered to be in a dark or 'off' state, and, above it, in a bright or 'on'
state it. Histograms of the durations of 'off' and 'on' state intervals were
produced, and fitted on a log scale to a linear polynomial using a
least-squares routine. Before fitting, the statistical weighting scheme of
refs.~\cite{ref:Crouch_Pelton_blinking,ref:kuno_JCP_2001} was applied to the
raw data, which produced the following probability distributions:

\begin{equation}
P(t_{off/on})=\frac{N(t_{off/on})}{N^{tot}(t_{off/on})}\frac{1}{\delta t_{avg, on/off}}.
\end{equation}

\begin{figure}[t]
\centerline{\includegraphics[width=8.5 cm]{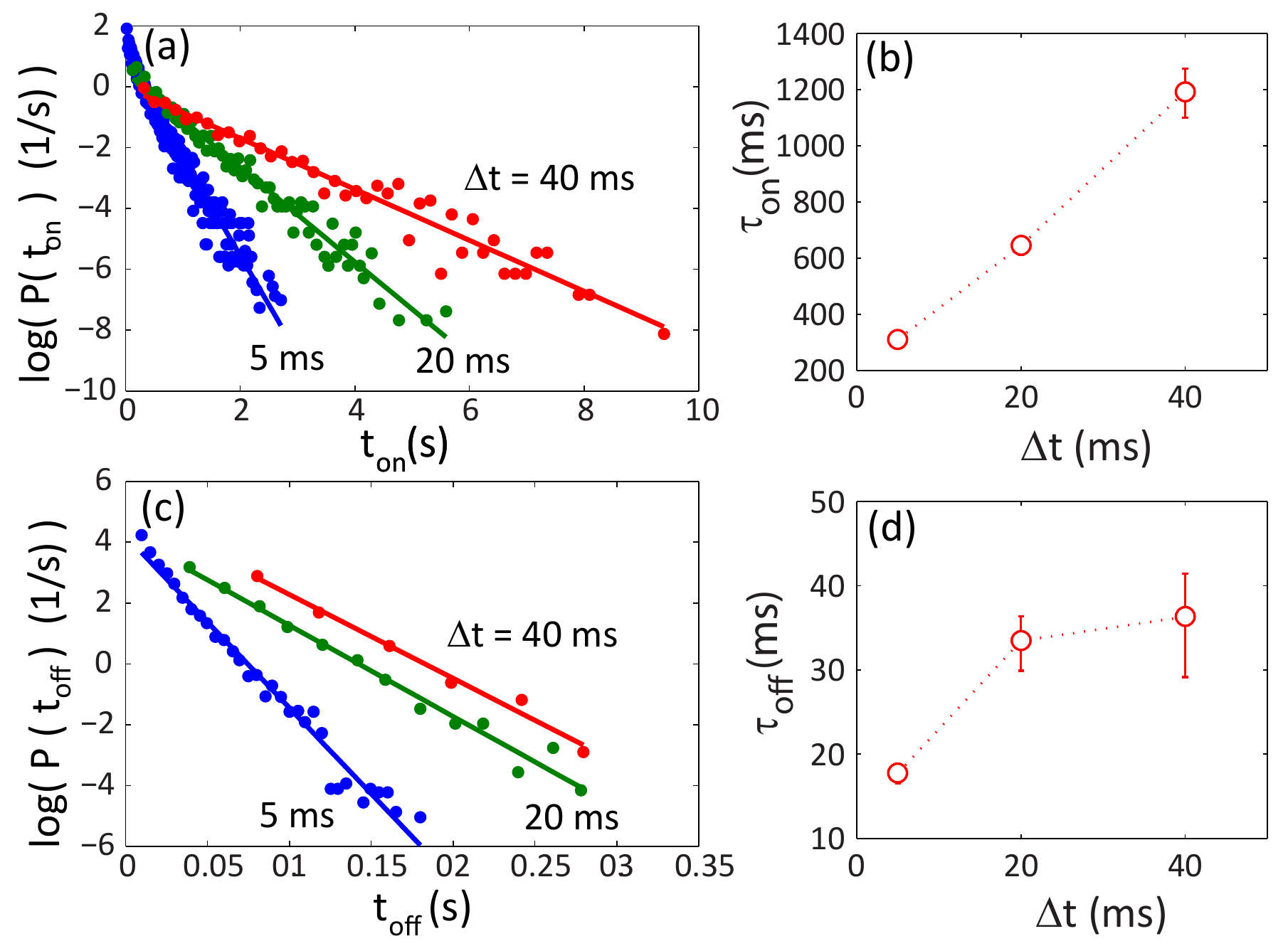}}
\caption{(a) Exponential fits (in log-linear scale) to weighted on-state time interval distributions, obtained for bin widths $\Delta t=$
5~ms, 20~ms and 40~ms, and intensity threshold $I_{th}=0.8I_{th,eq}$, with $I_{th,eq}$ as in eq.~(\ref{eq:Ith}). (b) Extracted characteristic times for on-states $t_{on}$. Error bars are $95~\%$
confidence intervals. (c) and (d) are the same for 'off'-times.}
\label{fig:SFig9}
\end{figure}

\begin{figure}[t]
\centerline{\includegraphics[width=8.5 cm]{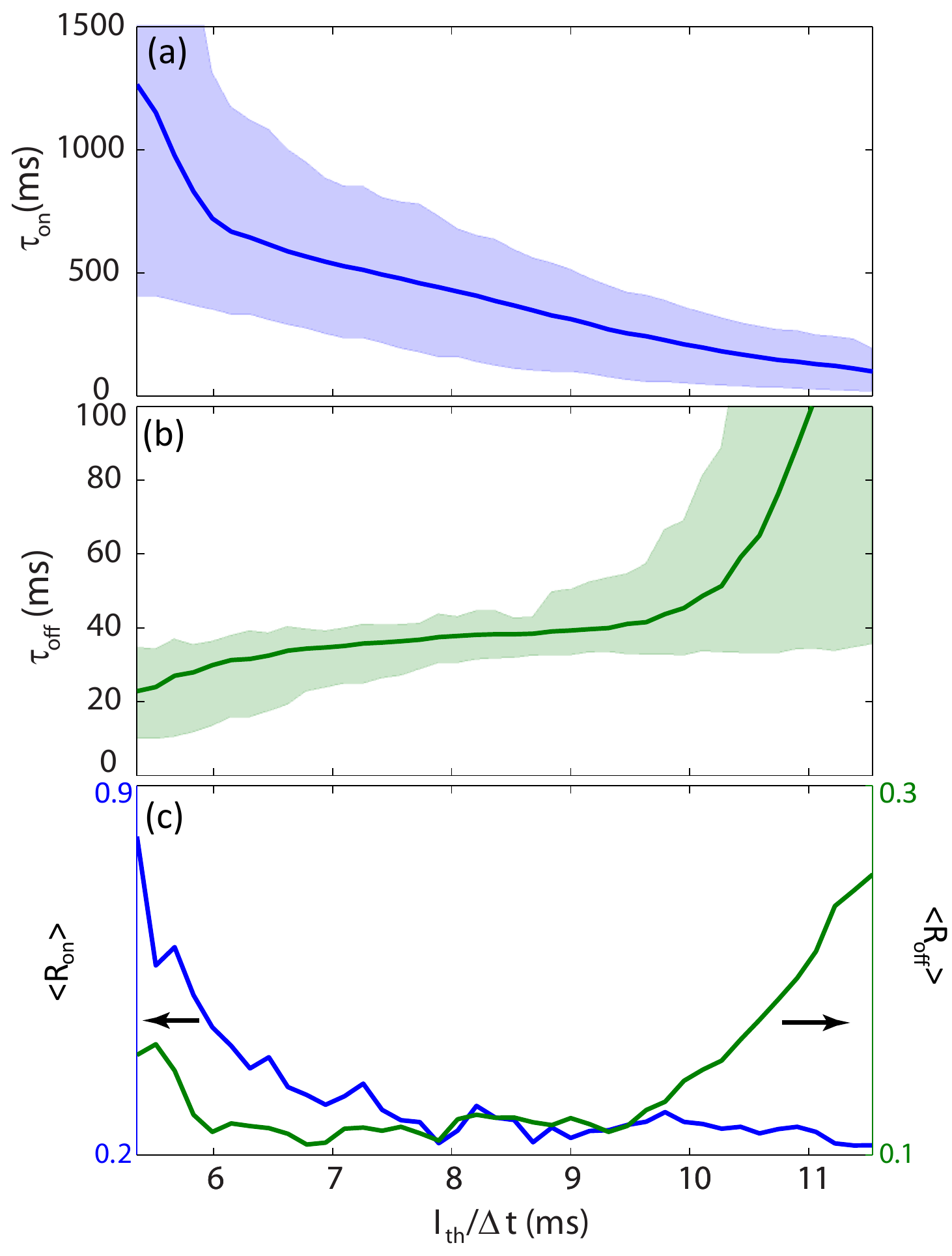}}
\caption{(a) and (b): Extracted characteristic times for 'on'- ($\tau_{on}$) and 'off'- ($\tau_{off}$)
time distributions, as functions of the threshold intensity $I_{th}$. Shaded areas show the spread of possible characteristic times extracted from data binned with $\Delta t$
varying between 5~ms and 40~ms. Thick lines are averages over time bins.(c). Linear fit figure of merit $R_{om/off}$ (eq.~(\ref{eq:FOM})) (averaged over all $\Delta t$)
 as a function of $I_{th}$.}
\label{fig:SFig10}
\end{figure}

Here, $N(t_{off/on})$ is the number of 'off' or 'on' events of duration
$t_{off/on}$, $N^{tot}(t_{off/on})$ is the total number of events in the time
series, and $\delta t_{avg, on/off}$ is the average time interval between
next neighbor events. This weighting is done to provide better statistical
estimates for the probability distributions at long times, where the data is
noisy and very sparse due to the finite duration of the measurement.
Figures~\ref{fig:SFig9}(a) and (b) respectively show 'on' and 'off' state
time interval distributions for $\Delta t=$ 5 ms, 20 ms and 40 ms, for an
intensity threshold $I=0.8\times I_{th,eq}$ indicated by a red dashed line in
Figs.~\ref{fig:SFig8}(a)-(c). Here, $I_{th,eq}$ is equidistant, in standard
deviations, from the 'on' and 'off' intensities $I_{off/on}$
~\cite{ref:Lippitz_Orrit_blinking_review}:
\begin{equation}
\label{eq:Ith}
\frac{I_{th,eq}-I_{off}}{\sqrt{I_{off}}}=\frac{I_{th,eq}-I_{on}}{\sqrt{I_{on}}},
\end{equation}
The closeness of the fits to the experimental data suggest that the 'on' and
'off' time probability distributions are single exponential, however it is
evident that the characteristic times differ by a significant amount for
different $\Delta t$ values. To understand how much data processing
influences the results, Figs.~\ref{fig:SFig10}(a) and (b) show the range of
characteristic times ($\tau_{on}$ and $\tau_{off}$) obtained, as a function
of $I_{th}$, for bin widths $\Delta t$ between 5 ms and 40 ms. The thick
lines in the figures correspond to averages over all $\Delta t$, at each
$I_{th}$.  The mean characteristic time for the 'on' state is seen to vary
widely, from close to 1330 ms for $I_{th}$ at the low intensity peak, to near
100 ms for $I_{th}$ at the high intensity peak, and a similar behavior is
observed for the 'off' times. Primarily, this large variation has to do with
the quality of the fits, which we evaluate through the 95~$\%$ confidence
intervals. For the single exponential functions $\exp(a_{on / off}\cdot t_{on
/ off}+b_{on / off})$, we plot, in Figs.~\ref{fig:SFig10}(c), the figure of
merit
\begin{equation}
\label{eq:FOM}
R_{on/off} = \sqrt{ \left(\frac{\delta a_{on/off}}{a_{on/off}}\right)^2+\left(\frac{\delta b_{on/off}}{b_{on/off}}\right)^2 },
\end{equation}
averaged over all bin widths $\Delta t$.

In~eq.(\ref{eq:FOM}), $\delta a_{on/off}$ and $\delta b_{on/off}$ are the
95~$\%$ fit confidence intervals. Evidently, the overall fit quality for 'on'
and 'off' times decreases significantly as $I_{th}$ approaches $I_{on}$ and
$I_{off}$ respectively, and reaches a baseline as $I_{th}$ moves towards the
opposite extremes. Due to the overlap between the bright and dark peaks in
the histograms of Figs.~\ref{fig:SFig8}(a)-(c), 'on' and 'off' times are
related through $I_{th}$, and so we may establish that reliable
characteristic times can be found only within $I_{th}$ intervals for which
$\left<R_{on/off}\right>$ are simultaneously minimized. Within the interval
7~ms$^-1< I_{th}/\Delta t<9$~ms$^{-1}$, $\tau_{off}$ ranges between 25 ms and
50 ms, and $\tau_{on}$ between 100~ms and 850~ms.

In summary, our analysis of the blinking trajectories shows that, within
valid threshold and bin size ranges, 'on' and 'off' time distributions can be
reasonably well-fit to single exponentials, as observed in other
self-assembled quantum dots; it is clear that the probability distributions
do not follow power laws as is the case for colloidal dots. On the other
hand, the uncertainties for the characteristic times are very wide, as
extracted values vary significantly with the choice of intensity threshold.
This limitation is strongly related to the poor resolution of the 'on' and
'off' peaks, which is indeed maximized over only a small range - at most a
decade - of bin widths.  An additional issue for the 'off' times distribution
is that the bin sizes over which peak resolution is clear are comparable to
the 'off' time characteristic times, while, ideally, $\tau_{off}/\Delta t
<10$ for good resolution of blinking events. As noted
in~\cite{ref:Crouch_Pelton_blinking}, it is apparent that blinking trajectory
analysis requires a very stringent set of conditions for the data, so that
reliable information may be extracted regarding the on/off time
distributions. This contrasts with the correlation analysis performed in the
main text, which is known to produce reliable quantitative information when
directly applied to raw data~\cite{ref:Lippitz_Orrit_blinking_review} -
albeit not directly providing information about the blinking interval
distributions.

%

%\bibliography{KS_bib_2013_12_17}

\end{document}